\documentclass[twocolumn,showpacs,preprintnumbers,amsmath,amssymb]{revtex4}
\usepackage{graphicx}% Include figure files
\usepackage{dcolumn}% Align table columns on decimal point
\usepackage{bm}% bold math

\begin{document}

\preprint{APS/123-QED}

\title{Half-metallic diluted antiferromagnetic semiconductors}

\author{H. Akai and M. Ogura}
\affiliation{Department of Physics, Graduate School of Science, Osaka University, 1-1 Machikaneyama, Toyonaka, Osaka 560-0043, Japan}

\date{\today}% It is always \today, today,

\begin{abstract}
The possibility of half-metallic antiferromagnetism, a special case of ferrimagnetism with a compensated magnetization, in the diluted magnetic semiconductors is highlighted on  the basis of the first principles electronic structure calculation. As typical examples, the   electrical and magnetic properties of II-VI compound semiconductors doped with $3d$    transition metal ion pairs---(V, Co) and (Fe, Cr)---are discussed.
\end{abstract}


\maketitle

Since the discovery of ferromagnetic diluted magnetic semiconductors (DMSs)\cite{mune, ohno, mats}, considerable attention has been paid to the origin and stability of their ferromagnetism\cite{akai, diet, jung, sato, sato2, kudr}. One of the main issues has been how to realize the magnetic transition temperature that exceeds room temperature for these DMSs. This is partly due to the possible applications of such materials to spin-electronics\cite{wolf}. Well-defined magnetization is certainly an indication of a  new degree of freedom that can store and transfer information. However, attaining  a sufficiently high magnetic transition temperature for such applications seems to be a  difficult task currently\cite{mats, sono, ploo, ando, yama}. Such a functionality that is expected for DMSs, however, is not  restricted only to the ferromagnetic DMSs. In this Letter, we propose a new candidate  for spin-electronics materials, namely, half-metallic diluted antiferromagnetic semiconductors. We expect that these materials will show significant spin-dependent electronic properties, and hence can be used as spin-electronics materials.

   The concept of half-metallic antiferromangets was first proposed by van Leuken and de Groot many years ago\cite{vanl}. They pointed out that in a special case of ferrimagnetism it might be possible for a system to become half-metallic, i.e., metallic in one spin direction of electrons but insulated in the other spin direction. Obviously, this is impossible for usual antiferromagnets because of the spin-rotational symmetry, but it is not necessarily the case for ferrimagnets with a compensated magnetic moment. Van Leuken and de Groot referred to such cases as half-metallic antiferromagnetism. Later, Pickett pointed out on the basis of first-principles calculation that La$_{2}$VCuO$_{6}$, La$_{2}$MnO$_{6}$, and La$_{2}$MnCoO$_{6}$ with a double-perovskite structure might be likely candidates for half metallic antiferromagnets\cite{pick}. However, thus far, the existence of such materials has not been verified experimentally. This indicates that intermetallic compounds or oxides may not be suitable candidates for half-metallic antiferromagnets. It could be possible that their magnetic and crystal structures are not as stable energetically as predicted by LDA (local density approximation) calculations.

   In contrast to the above cases where the band structure is hardly controlled by small modifications of the component atoms and/or the crystal geometries, doped semiconductors provide us with a much wider range of choices to obtain half-metallic antiferromagnets. This is because they often form impurity bands within a band gap. Due to the existence of the gap in the host semiconductors, these impurity bands sometimes also have a gap at the Fermi level. In this regard, as long as the impurity ions carry a local magnetic moment, as in the case of DMSs, the system can easily become half-metallic. 

\begin{figure}
\includegraphics[width=19pc]{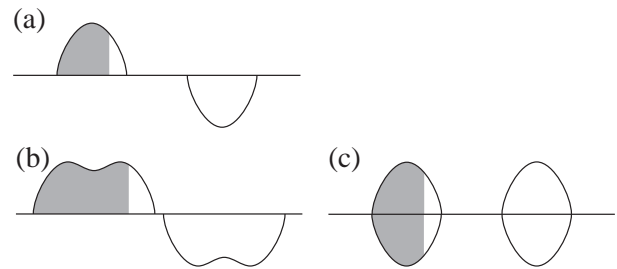}
\caption{\label{fig1} Schematic views of the electronic structure of Cr. The local density of the Cr state is shown as a function of the energy. The shaded part indicates the occupied state. (a) Single Cr ion embedded in the host.  (b) Two Cr ions with a ferromagnetic coupling.  (c) Two Cr ions with an antiferromagnetic coupling.}
\end{figure}

   First let us observe the electronic structure of ferromagnetic DMSs briefly. Consider II-VI compound semiconductors doped with $3d$ magnetic ions, such as (Zn, Cr)S. In this case, well-defined impurity bands originating from Cr are formed within the band gap. Irrespective of the magnetic structure, a local magnetic moment of approximately 4 $\mu_{\rm B}$ is formed on the Cr ions. The electronic structure of a single Cr ion in the host is schematically shown in Fig.~\ref{fig1} (a). If two Cr ions with such an electronic structure forms a ferromagnetic coupling, the resulting electronic structure of the system is as shown in Fig.~\ref{fig1} (b). On the other hand, if it forms an antiferromagnetic coupling, the electronic structure appears different as shown in Fig.~\ref{fig1} (c). The ferromagnetic coupling realizes a half-metallic structure although the antiferromagnetic coupling does not. In these cases, the energy gain due to the coupling is much larger for the ferromagnetic coupling, which is referred to as double exchange, than the superexchange of the antiferromangetic coupling, as pointed out by one of the present authors\cite{akai}. Therefore, such systems become a half-metallic ferromagnet in their ground state.

\begin{figure}
\includegraphics[width=19pc]{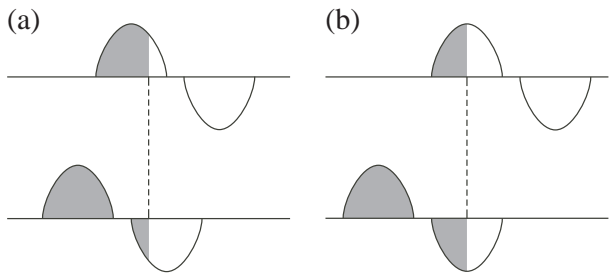}
\caption{\label{fig2} (a) Schematic views of the electronic structure (density of states) of Cr (upper) and Fe (lower).  (b) The case where each spin band is half-filled.}
\end{figure}

\begin{figure}
\includegraphics[width=9pc]{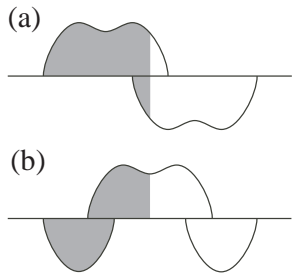}
\caption{\label{fig3} Schematic views of the electronic structure (density of states) of Cr and Fe.  (a) Cr and Fe with a ferromagnetic coupling.  (b) Cr and Fe with an antiferromagnetic coupling.}
\end{figure}

   We can make similar arguments for the half-metallic antiferromagnetic DMSs. It is obvious from a symmetry consideration that we need at least two different types of magnetic ions to realize a half-metallic structure. We consider two $3d$ magnetic ions: one of them has less than half-filled $d$-electrons, for example Cr, while the other has more than half-filled $d$-electrons, for exmaple Fe. When the host is a II-VI compound semiconductor, a Cr ion is usually in the nominally $d^{4}$ configuration and Fe is in the nominally $d^6$ configuration. In this case, the electronic structures of a single Cr and Fe ion are as shown in Fig.~\ref{fig2} (a). The results of the coupling between the Cr and Fe ions, which can either be ferromagnetic or antiferromagnetic, are schematically shown in Fig.~\ref{fig3} (a) and (b). Though the ferromagnetic coupling produces a metallic band in both spin directions, the antiferromagnetic coupling realizes the half-metallic bands.

   With regard to the gain in the band energy, which is relevant to the coupling between ions, the band energy for antiferromagnetic coupling is larger than that of the ferromagnetic coupling. This is because the antiferromagnetic coupling produces a common band for Cr and Fe in the up spin state (we choose the metallic band as up spin). This is a basic requirement for the double exchange to occur because the double exchange is simply the hybridization between two degenerate (or nearly degenerate) local states. This situation would be ideal if each spin band was half-filled as shown in Fig.~\ref{fig2} (b). On the other hand, the antiferromagnetic coupling causes hybridization between energetically separated states. We may call the latter as a superexchange though it is actually ferromagnetic as opposed to the usual superexchange that is mostly the origin of antiferromagnetism. Using the same argument as that for ferromagnetic DMSs\cite{akai}, the double exchange should be much stronger than the superexchange, thus causing antiferromagnetic ordering in the present case.
   
\begin{figure}
\includegraphics[width=6pc]{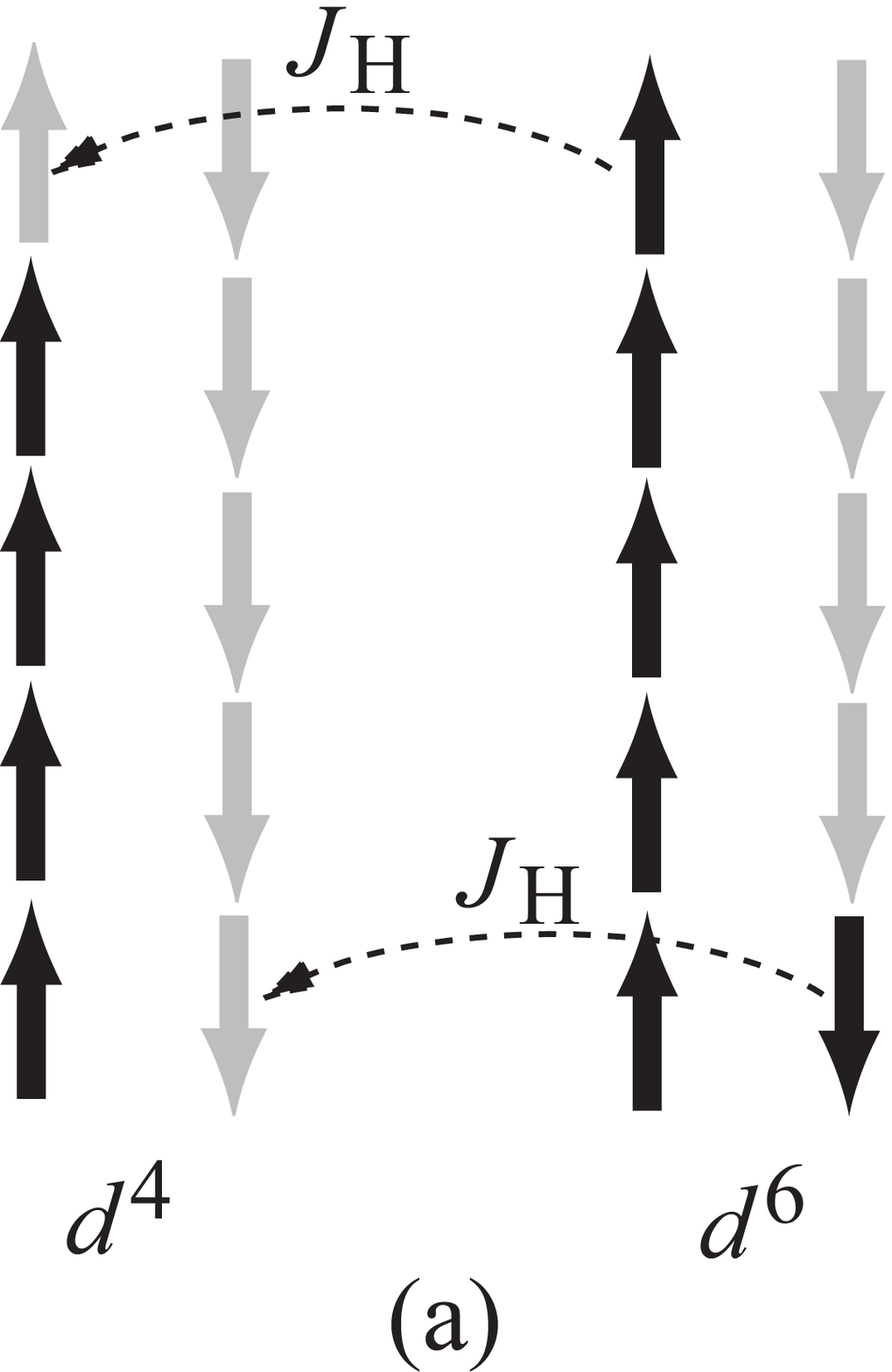}
\hskip 3pc
\includegraphics[width=6pc]{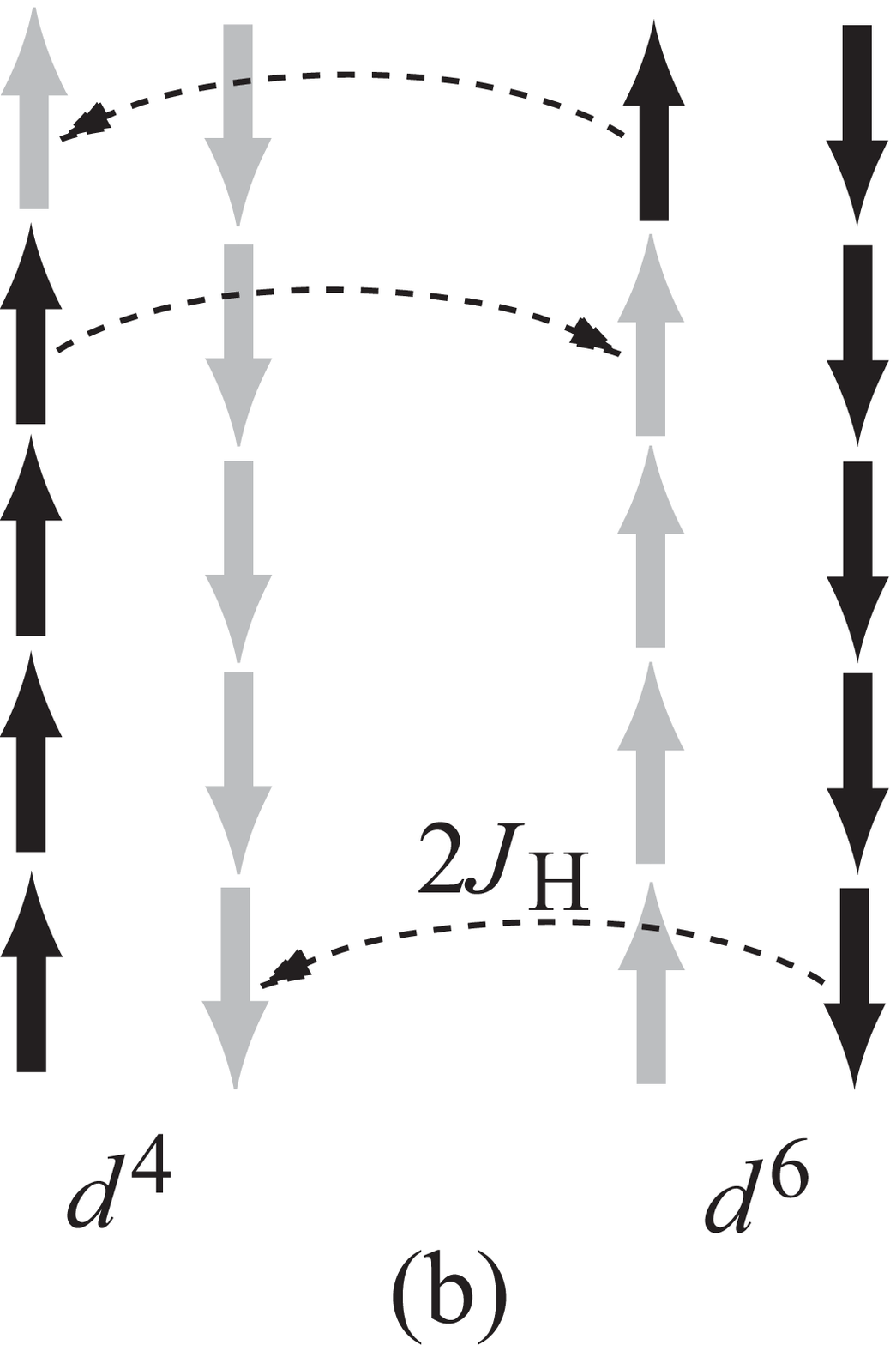}
\caption{\label{fig4} A pair of the $d^4$ and $d^6$ ions with (a) ferromagnetic coupling and (b) antiferromagnetic coupling. The black and gray arrows indicate occupied and unoccupied electron states, respectively, and the dotted arrows indicate possible electron transfer.}
\end{figure}

   In the above, the mechanism that stabilizes an antiferromagnetic coupling was explained in the framework of band structure. The same mechanism can also be described in terms of many-electron states in an analogous way to that for ferromagnetic double exchange. Let us consider a pair of $d^4$ and $d^6$ ions. When it forms a ferromagnetic coupling (Fig.~\ref{fig4} (a)), there are two possible ways to transfer an electron from one ion to another: to move one of up-spin electrons from the $d^6$ to $d^4$ ion or to move one of down-spin electrons from the $d^6$ to $d^4$ ion. Both ways, however,  violate Hund's first law and each requires an additional energy of $\sim J_{\rm H}$ (intra-atomic exchange energy). In the case of antiferromagnetic coupling (Fig.~\ref{fig4} (b)), in addition to the above two possibilities, it is also possible to move a down-spin electron from the $d^4$ to $d^6$ ion. Whereas the transfer of an up-spin electron from the $d^6$ to $d^4$ ion violates Hund's law twice (first at the $d^6$ ion and second at the $d^4$ ion), other transfers does not. Since the electron transfers without an additional energy lowered the total energy of an ion pair to the first order of the transfer integral, the antiferromagnetic coupling energetically more stable than the ferromagnetic one.

   In order to verify the above concept, we have performed the first-principles electronic structure calculation by the KKR-CPA-LDA method (Korringa-Kohn-Rostoker GreenÕs function method with coherent potential approximation and local density approximation of the density functional method). The details of the calculation are mostly the same as those used in our previous paper\cite{akai} and used by other authors for similar calculations on ferromagnetic DMSs\cite{sato}. The only difference is that we introduce two magnetic ions instead of one: these ions act as a substitute for the host atoms.

\begin{figure}
\includegraphics[width=15pc]{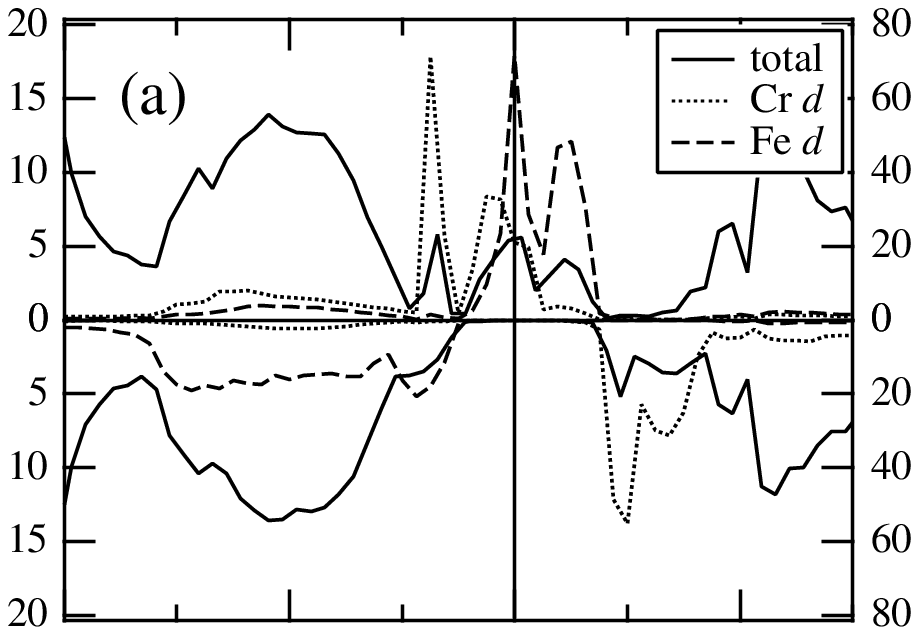}
\includegraphics[width=15pc]{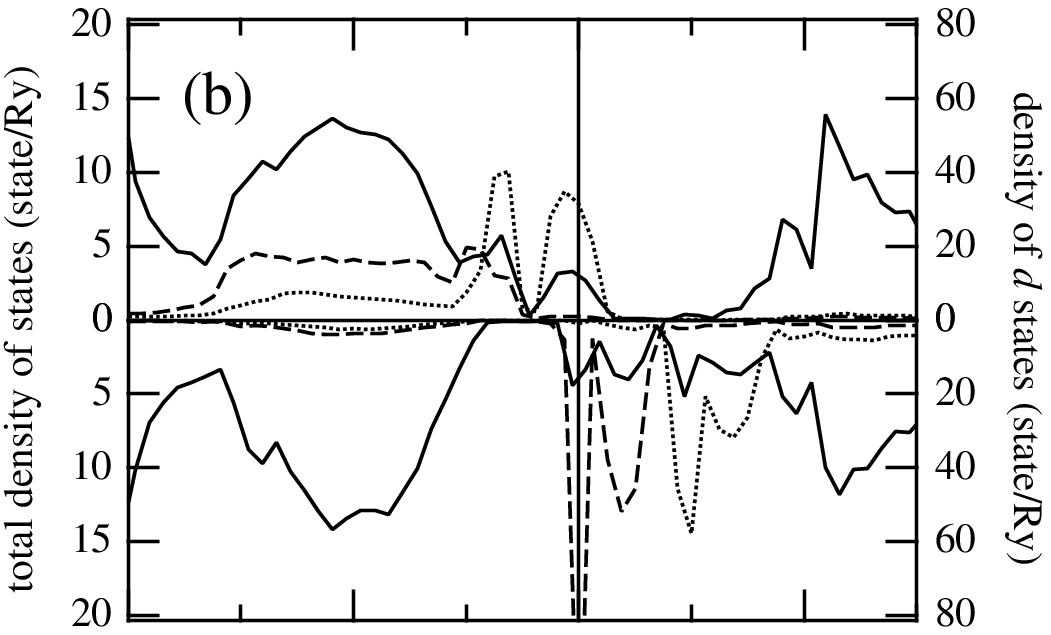}
\includegraphics[width=15pc]{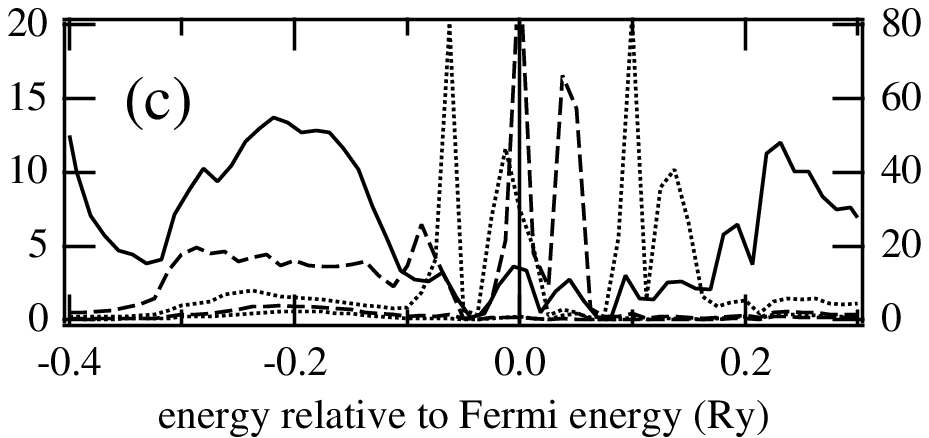}
\caption{\label{crfe} Density of states (DOS) for (Zn$_{0.9}$Cr$_{0.05}$Fe$_{0.05}$)S.  (a) DOS for the antiferromagnetic state.  The solid line shows the total DOS (left axis) and the dashed lines show the local $d$ DOS of Cr and Fe (right axis).  The DOS for the up and down spin are drawn in the upper and lower half of the panel, respectively.  (b) Ferromagnetic state and (c) LMD state.}
\end{figure}

   Figure~\ref{crfe} shows the calculated electronic structure of (Zn$_{0.9}$Cr$_{0.05}$Fe$_{0.05}$)S. The three cases are as follows: (a) is for the antiferromagnetic state; (b) the ferromagnetic state; and (c) the local-moment-disordered (LMD) state that simulates a paramagmetic or spinglass state. It is apparent  that a half-metallic electronic structure can be realized only by the antiferromagnetic state. The magnetization totally vanishes in the antiferromagnetic case. Though the magnetization is not necessarily zero when considering the symmetry, the existence of a gap for down spin state (the definition of the direction of the spin is the same as mentioned earlier) prevents each spin state to take an arbitrary occupation number. In the case of doped II-VI compound semiconductors, it is likely that the magnetic moment per impurity atom assumes an integer value. For the ($d^{4}, d^{6}$) pair, which is the present case, the vanishing magnetization for the antiferromagnetic case seems reasonable.
      
   The total energy for the antiferromagnetic and ferromagnetic states are $-0.418$ mRy and 0.182 mRy, respectively, compared with the LMD state; this indicates that the antiferromagnetic state should be the ground state. The energy difference between the antiferromagnetic and LMD states gives a fair estimation of the N\'{e}el temperature if suitable factors are included, i.e.,  divided by the concentration of the magnetic ions and multiplied by $2 \over 3$. In the present case of (Zn$_{0.9}$Cr$_{0.05}$Fe$_{0.05}$)S, the N\'{e}el temperature is 440 K. The real N\'{e}el temperature, however, could be considerably lower than 440 K due to the fluctuations in the spin state and that in the atomic arrangement (configuration)\cite{sato}.

   The local density of states indicates that our schematic image clearly reveals the structure of the impurity bands formed by the two different magnetic ions; this is well understood in terms of the hybridization between the local electron states originating from those magnetic ions. The essential point is to lower the band energy, i.e., the sum of the single particle energy, through the hybridization, which determines the nature of the coupling.
   
\begin{figure}
\includegraphics[width=15pc]{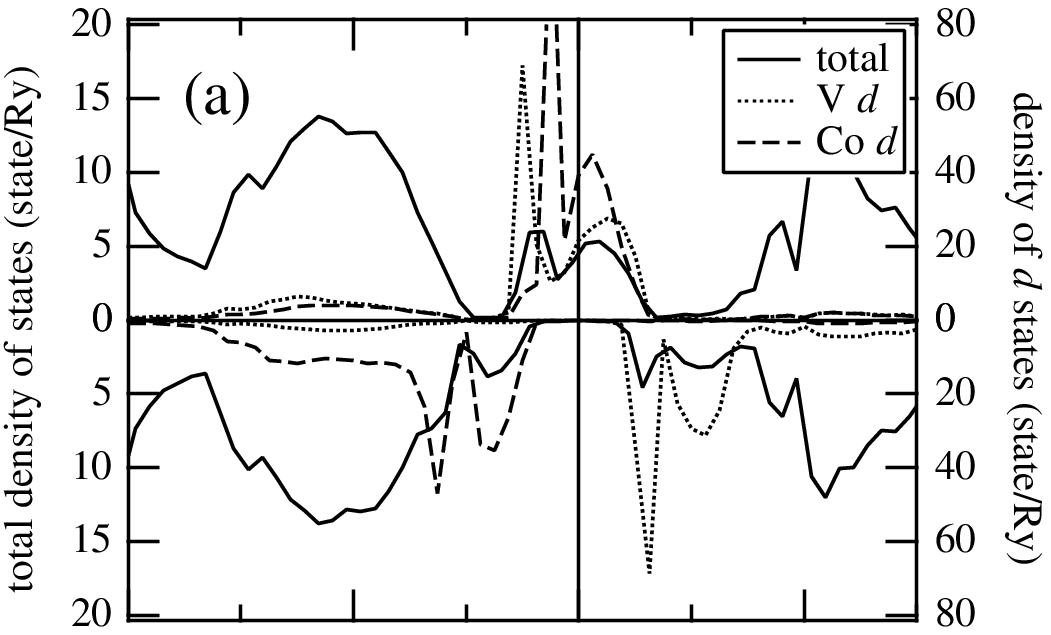}
\includegraphics[width=15pc]{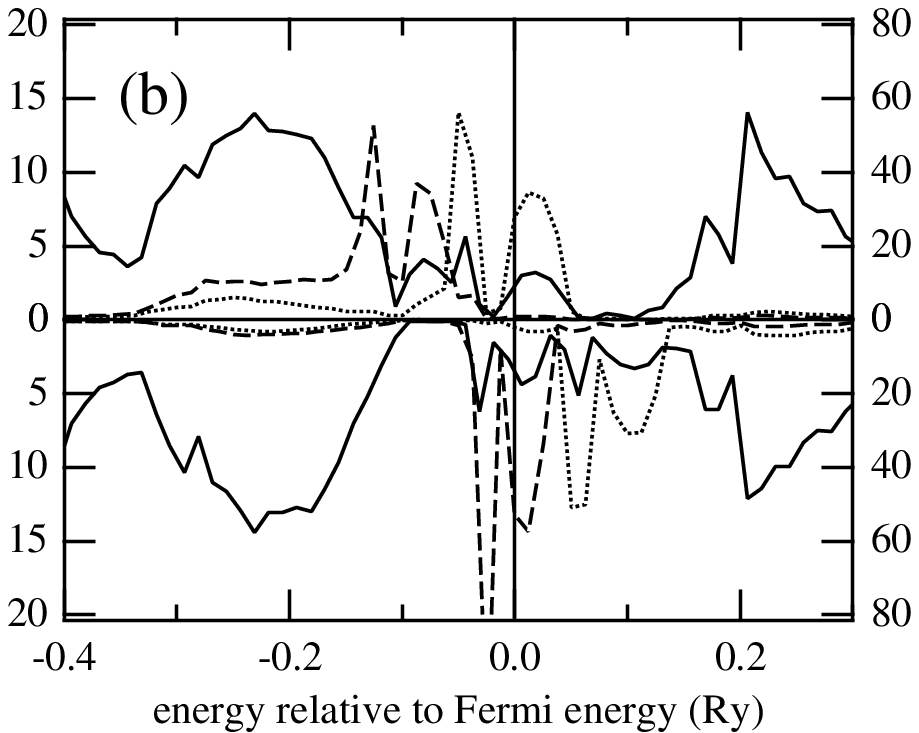}
\caption{\label{vco} DOS of (Zn$_{0.9}$V$_{0.05}$Co$_{0.05}$)S.}
\end{figure}

\begin{figure}
\includegraphics[width=16pc]{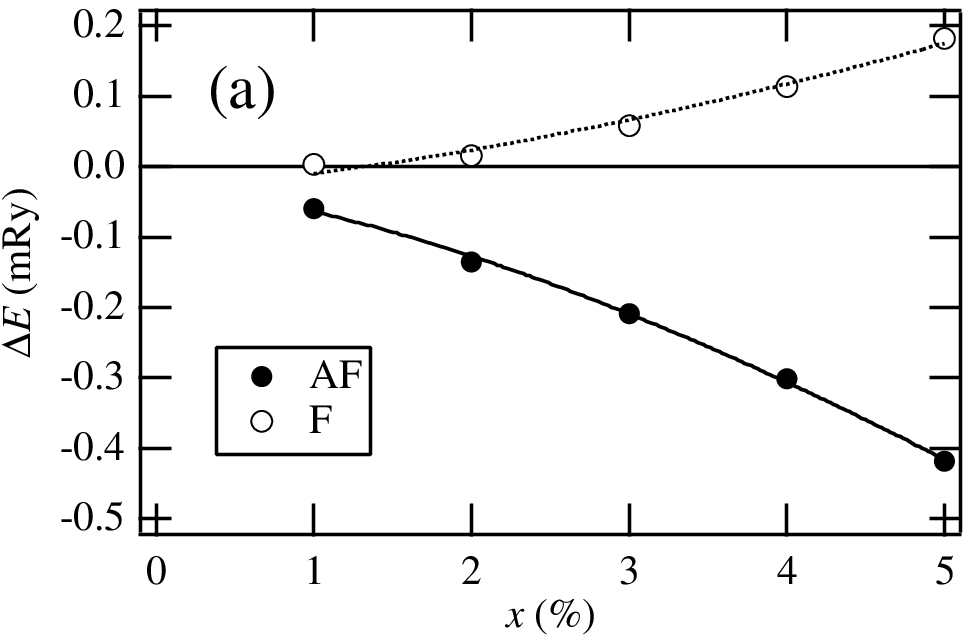}
\includegraphics[width=16pc]{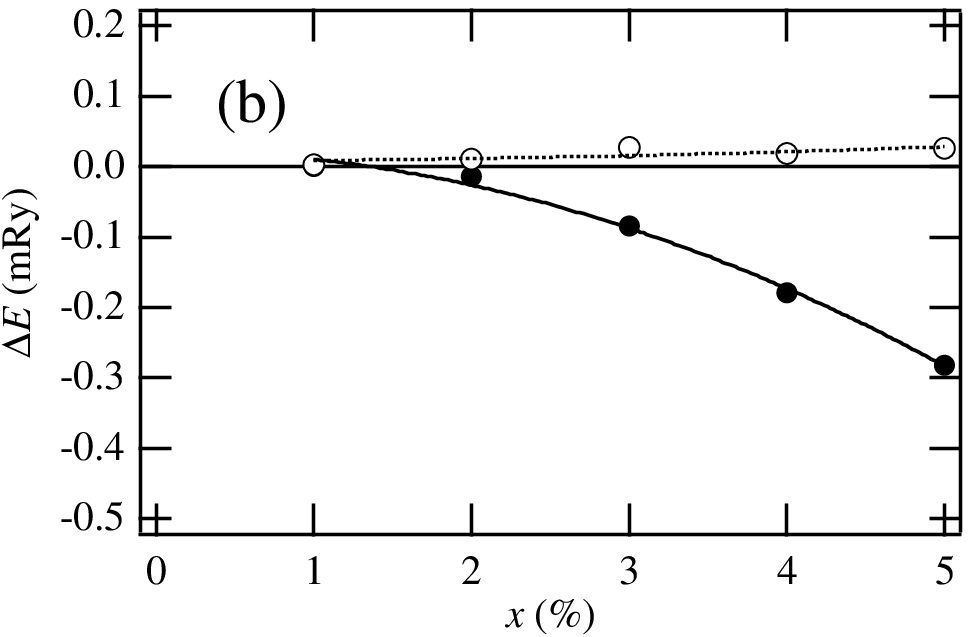}
\caption{\label{energy} (a) Total energy of (Zn$_{1-2x}$Cr$_{x}$Fe$_{x}$)S.  The horizontal axis represents the impurity concentration $x$.  The closed and open circles correspond to the antiferromagnetic and ferromagnetic state, respectively.  (b) Total energy of (Zn$_{1-2x}$V$_{x}$Co$_{x}$)S.}
\end{figure}

   Figure~\ref{vco} shows a similar plot for (Zn$_{0.9}$V$_{0.05}$Co$_{0.05}$)S. This is the case of the $d^{3}$ and $d^{7}$ combination of the magnetic ions. Again, the antiferromagnetic state is half-metallic and energetically the most stable. The magnetization in the antiferromagnetic case is again exactly zero. The overall behavior of the DOS is similar to (Zn$_{0.9}$Cr$_{0.05}$Fe$_{0.05}$)S, but in this case, the feature of the common bands for the up spin state and that of the split band for the down spin state are more clearly observed. Compared with the LMD state, the total energy in this case are $-0.282$ mRy and 0.026 mRy for the antiferromagnetic and ferromagnetic state, respectively, thus, the expected N\'{e}el temperature is lower than that in the case of (Zn$_{0.9}$Cr$_{0.05}$Fe$_{0.05}$)S. Since each spin state of a single ion (i.e., before hybridization between the magnetic ions) with regard to (Zn$_{0.9}$V$_{0.05}$Co$_{0.05}$)S is closer to half-filling than (Zn$_{0.9}$Cr$_{0.05}$Fe$_{0.05}$)S, a stronger double exchange can be expected for the former as opposed to the result of the calculation. This indicates that the local magnetic moments of V and Co are considerably smaller than those of Fe and Cr. This causes a smaller exchange splitting, thus reducing the energy gain due to the double exchange as compared to the superexchange.
     
   The concentration dependence of the total energy is also shown in Fig.~\ref{energy}. The behavior near the impurity limit is well fitted by $\propto x^{3 \over 2}$ for (Zn, Cr, Fe)S, indicating that the double exchange plays a role in determining the magnetic coupling. On the other hand, the concentration dependence for (Zn, V, Co)S is expressed as $\propto x^2$. This implies that the hybridization of the $d$ states of magnetic ions to the $p$ states of the surrounding S atoms also contributes to the magnetic coupling\cite{sato}.

   In order to obtain a compensated magnetic moment in such a half-metallic antiferromagnet (or ferrimagnet), it is essential that the sum of valence $d$-electron numbers of two (or more) types of ions is 10, such as ($d^{3}, d^{7}$) and ($d^{4}, d^{6}$). In the case of II-VI compounds semiconductors, this combination is (V, Co) and (Cr, Fe). On the other hand, in the case of III-V, this combination should be (Cr, Ni) and (Mn, Co). For the other combinations, it is still possible to obtain a half-metallic ferrimagnet but the magnetization in such cases would not be completely compensated.
   
   In summary, we have proposed a new magnetic system composed of semiconductors doped with more than two types of magnetic ions; the sum of their valence $d$-electrons is 10. We predict that significatn feature of these systems to be half-metallic antiferromagnetism (or the special case of ferrimagnetism with compensated magnetic moments.) Since the systems have 100\% spin polarized Fermi surfaces like usual ferromagnetic DMSs, they can also be used as spin-electronics materials.

\begin{acknowledgments}
This work is partly supported by Special Coordination Funds for the Promotion of Science and Technology, Leading Research ``Nanospintronics Design and Realization'', by Nano Technology Program, New Energy and Industrial Technology Development Organization (NEDO), by 
MEXT KAKENHI, No.\ 16031210, 17064008, and by JSPS KAKENHI No.\ 17340095.
We are very grateful to Prof.\ Dr.\  P.H.\  Dederichs in Forschungszentrum J{\" u}lich for valuable discussion. 
\end{acknowledgments}

\end{document}